\documentclass[10pt,nofootinbib,amsmath]{revtex4}
\usepackage{graphicx}
\usepackage{bbold}
\usepackage{bbm}
\newcommand{\be}{\nopagebreak[3]\begin{equation}}
\newcommand{\ee}{\end{equation}}
\newcommand{\ba}{\nopagebreak[3]\begin{eqnarray}}
\newcommand{\ea}{\end{eqnarray}}

\begin{document}

\title{Self-energy and vertex radiative corrections in LQG}
\author{Claudio Perini$^{ab}$, Carlo Rovelli$^a$, Simone Speziale$^{ac}$}

\affiliation{\small\it ${}^a$Centre de Physique Th\'eorique de Luminy\footnote{Unit\'e mixte de recherche (UMR 6207) du CNRS et des Universit\'es de Provence (Aix-Marseille I), de la M\'editerran\'ee (Aix-Marseille II) et du Sud (Toulon-Var); laboratoire affili\'e \`a la FRUMAM (FR 2291).}, Case 907, F-13288 Marseille, EU \\ 
\small\it ${}^b$Dipartimento di Matematica, Universit\`a Roma Tre, Largo Murialdo 1,  I-00146 Roma, EU\\
\small\it ${}^c$Perimeter Institute, 31 Caroline St North, Waterloo, ON N2L 2Y5, Canada}
\date{\small\today}
{\abstract \noindent We consider the elementary radiative-correction terms in loop quantum gravity.  These are a two-vertex ``elementary bubble" and a five-vertex ``ball"; they correspond to the one-loop self-energy and the one-loop vertex correction of ordinary quantum field theory. We compute their naive degree of (infrared) divergence.}
\maketitle

\vskip1cm

\section{Introduction}

\noindent Recent developments in the spinfoam research program \cite{EPR,EPR2,LS,FK,EPRL} have provided a novel and manageable perturbative definition of the dynamics of 
loop quantum gravity \cite{book,lqg2}.   The theory is first cut-off by choosing a 4d triangulation 
$\Delta_N$ of spacetime, formed by $N$ 4-simplices; then the continuous theory can be defined by the $N\to\infty$ limit of the expectation values.\footnote{This is similar to the well-known nonperturbative definition of QCD via a lattice  \cite{Wilson}, but with the difference that  because of general covariance the lattice spacing does not affect physical expectation values.}  At fixed $N$, the partition function of the theory is given by a sum over spins and intertwiners, which can be interpreted as a version of the Misner-Hawking ``sum over geometries" \cite{misner}, regulated by replacing continuous geometries with Regge geometries \cite{Regge} on  $\Delta_N$. This sum may contain divergent terms.  Here we study these terms.  For previous studies of divergences in spinfoam models see \cite{Perez:2000bf,AC,C1,C2}. 

We consider the version of the formalism defined in \cite{EPRL}, which coincides
with the one discussed in \cite{FK} when the Immirzi parameter $\gamma\in I\!\!R^+$ satisfies $\gamma<1$.  The theory is defined by the partition function
\be
Z=\sum_{j_f,i_e} \prod_f\;d(j_f)\;\prod_v\;A_v(j_f,i_e).
\label{partition}
\ee
The sum is over $SU(2)$ irreducible representations $j_f$ associated to the triangles $f$ of $\Delta_N$ and $SU(2)$ intertwiners $i_e$ associated to tetrahedra $e$ of $\Delta_N$. 
$d(j)$ is a face amplitude and the vertex amplitude is given by 
\be
A_v(j_{f},i_e) \equiv  \lambda \sum_{i_e^{+},i^-_e} 15j_N\!\!\left(j^+_{f}, i_e^+\right) 
\  15j_N\!\!\left(j^-_{f},  i_e^-\right) \prod_e f_{i_e^+,i^-_e}^{i_e}, 
\label{vertex}
\ee
\vskip-2mm\noindent
where $j^{\scriptscriptstyle \pm}\equiv\frac{|1\pm\gamma|}{2}j$;\  $15j_N$ denotes the contraction of five normalized 4-valent intertwiners according to the pattern of a 4-simplex (see \eqref{15jN}), and $f_{i^+,i^-}^i$ are the fusion coefficients \cite{EPR}, discussed in detail below. The arguments of $A_v$ are the ten spin and the five intertwiners (labeled by virtual spins in a coupling channel)  adjacent to the vertex $v$.  We have inserted a dimensionless coupling constant $\lambda$. This vertex amplitude is derived in \cite{EPR,EPR2,LS,FK,EPRL} from the action of GR; see these references for details. On the semiclassical behavior of the sum (\ref{partition}), see \cite{semiclassical}.  

Transition amplitudes can be constructed by picking a triangulation $\Delta_N$ with a boundary $\partial\Delta_N$, and fixing spins and edges on the boundary in (\ref{partition}): 
\be
W(j_l,i_n)=\sum_{j_f,i_e} \prod_f\;d(j_f)\;\prod_v\;A(j_f,i_e),
\label{amplitudes}
\ee
\vskip-1mm\noindent
where $j_l$ are the spins of the boundary faces (or ``links") and $i_n$ are the intertwiners of the boundary tetrahedra (or ``nodes"). (The expressions ``links" and ``nodes" refer to the graph $\Gamma$ defined by the 1-skeleton of the cellular complex dual to  $\partial\Delta_N$.) We are interested in the potential divergences of the sums (\ref{amplitudes}). 

As explained in the next Section, the structure of such divergences bear some similarities (but also some differences) with that of the divergences in Feynman diagrams.  There, divergences come from momentum integrals associated to internal \emph{loops}. Here, divergences turn out to be associated to \emph{bubbles} in the triangulation. A \emph{bubble} is a collection of faces in the 2-skeleton of the cellular complex dual to $\Delta_N$, forming a closed 2-surface.  As a first step towards the characterization of all divergences, we study here the most elementary of such bubbles, and the potential divergences they give rise to.  In particular, we focus on two elementary graphs: the self-energy (or ``elementary bubble"), and the vertex divergence (or ``elementary ball").   As in QFT, these can be viewed as divergences on particularly simple triangulations with boundaries (a $\Delta_2$ and a $\Delta_5$), or as divergences arising in sub-triangulations in larger triangulations. 

It is shown in \cite{Reisenberger:2000zc} that any spinfoam model can be expressed as a Feynman amplitude of a group field theory  \cite{GFT}.  If we view (\ref{amplitudes}) as determined by the Feynman expansion of a group field theory, these graphs correspond precisely to the self-energy correction to the propagator and the first radiative correction of the vertex, from which, in several QFT's, all other divergences depend.

Our results are summarized below, in Section \ref{results}.  We find that the naive degree of divergence depends crucially on the details of the normalization of the fusion coefficients mentioned above, and the faces amplitude.  It is therefore important to gain a better control over these quantities, which are not clearly fixed in the literature. For an appropriate and natural choice of the normalization of the fusion coefficients, both diagrams are finite. 

\vskip1cm

\section{Preliminaries}

\subsection{Bubbles}

At fixed spins $j_f$, the sum over the intertwiners $i_e$ in  (\ref{amplitudes})  is always over a finite number of terms.  Therefore divergences can only arise from $j_f\to\infty$.   The spins $j_f$ are constrained by the Clebsch-Gordan inequalities at each edge, because when these relations are not satisfied the vertex amplitude vanishes. This fact determines the structure of the possible divergences \cite{Perez:2000bf,AC}. 

A well-known way of resolving the Clebsch-Gordan inequalities, indeed, is to change variables as follows. Replace a face with spin $j_f$ by $2j_f$ overlapping sheets. At a given edge $e$, the Clebsch-Gordan conditions are satisfied if and only if there is a way of pairing all individual sheets coming from the four faces that join at the edge, in such a way that each sheet is paired with one and only one sheet of a different face.  The pairing joins individual sheets across faces, forming surfaces that wrap over the 2-skeleton of the triangulation.   Such surfaces can be of two kinds: either they meet the boundary, or they are closed. Call the closed surfaces ``bubbles".   Suppose we want to increase the spins $j_f$ to infinity (keeping the boundary spins $j_l$ fixed).  Then we have to add more and more of such surfaces.  However, we cannot add surfaces that meet the boundary, because this would increase $j_l$.  Therefore the only possibility is to increase the number of surfaces of the closed kind, or ``bubbles" \cite{Perez:2000bf,AC}. 

The situation is thus similar to that in conventional QFT, where momentum conservation at QFT vertices implies that divergences are necessarily associated with closed lines on the Feynman graph, or ``loops".  Here Clebsch-Gordan inequalities at each edge imply that divergences are necessarily associated with ``bubbles". 

\subsection{Normalizations}

The convergence properties of (\ref{partition}) depend on both face and vertex amplitudes.  The recent developments in the spinfoam dynamics have resulted in a vertex amplitude that, at least for $\gamma<1$, is unique up to normalization; while the face amplitudes 
are not fully fixed yet. At the present stage of investigation, it is therefore 
useful to study the convergence properties of a family of models like (\ref{partition}), which 
differ by the face amplitudes and the normalization of the vertex.

The face amplitude $d(j)$ in (\ref{partition}) was taken in \cite{EPRL} to be the dimension of the $SO(4)$ representation determined (via the Plebanski constraints) by the $SU(2)$ representation $j$
\be
d(j)= d_{SO(4)}(j) \equiv (2j^-\!\!+1)(2j^+\!\!+1)=  \left(|1-\gamma|j+1\right)\left((1+\gamma)j+1\right)
\label{SO(4)}
\ee
on the basis of an analogy with BF theory.  However, this part of the definition of the theory is the most uncertain.  A natural alternative is to chose the dimension of the representation $j$ itself
\be
d(j) = d_{SU(2)}(j) \equiv \left(2j+1\right).
\label{SU(2)}
\ee
To keep the issue open, we consider here a generic dependence such that 
\be
d(j) \sim j^k
\ee
when $j\to\infty$. The two cases considered above correspond to $k=2$ and $k=1$, respectively. 

The fusion coefficients entering (\ref{vertex}) are covariant maps from the $SO(4)$ intertwiner space to the $SO(4)$ intertwiner space. They were defined in \cite{EPR} as
\be
   f^i_{i^+ i^-}(j_a)=\langle i^+ i^-|f|i \rangle =
   i^{abcd}\ C_a^{a^+a^-}C_b^{b^+b^-}C_c^{c^+c^-}C_d^{d^+d^-}\
   i^+_{a^+b^+c^+d^+}\ 
   i^-_{a^-b^-c^-d^-}. 
   \label{fusion}
\ee
where $i^{abcd}$ are normalized 4-valent intertwiners and $C_a^{a^+a^-}= \langle j_1^+j_1^-, a^+a^-|j_1a \rangle$ are Clebsch-Gordan coefficients. See Appendix A for precise definitions.
(The group integral appearing in the definition in \cite{EPR} is not needed once we project on the $SO(4)$ invariant subspace spanned by the $|i^+ i^- \rangle$ intertwiners.)
It is not clear to us what are the conditions that fix the normalization of the fusion coefficients defined in \cite{EPR}.    In particular, there is a more natural alternative given by
\be
   \tilde f^i_{i^+ i^-}(j_a)=\langle i^+ i^-|f|i \rangle =
   i^{abcd}\ i_a^{a^+a^-}i_b^{b^+b^-}i_c^{c^+c^-}i_d^{d^+d^-}\
   i^+_{a^+b^+c^+d^+}\ 
   i^-_{a^-b^-c^-d^-}
   \label{fusion2}
\ee
where $i_a^{bc}$ are normalized trivalent intertwiners (see Appendix); these fusion coefficients are related to the above ones by
\be\label{pippo}
    f^i_{i^+ i^-}(j_a)=\sqrt{\prod_{a=1...4}(2j_a+1)}\ \ \tilde f^i_{i^+ i^-}(j_a).
   \ee
This uncertainty in the normalizations is similar to the one in old spinfoam models: see  \cite{Perez:2000bf,AC,C1,C2} and \cite{book}.

\vskip1cm


\section{The elementary bubble}

\begin{figure}[b]
\centering
\includegraphics[scale=0.12]{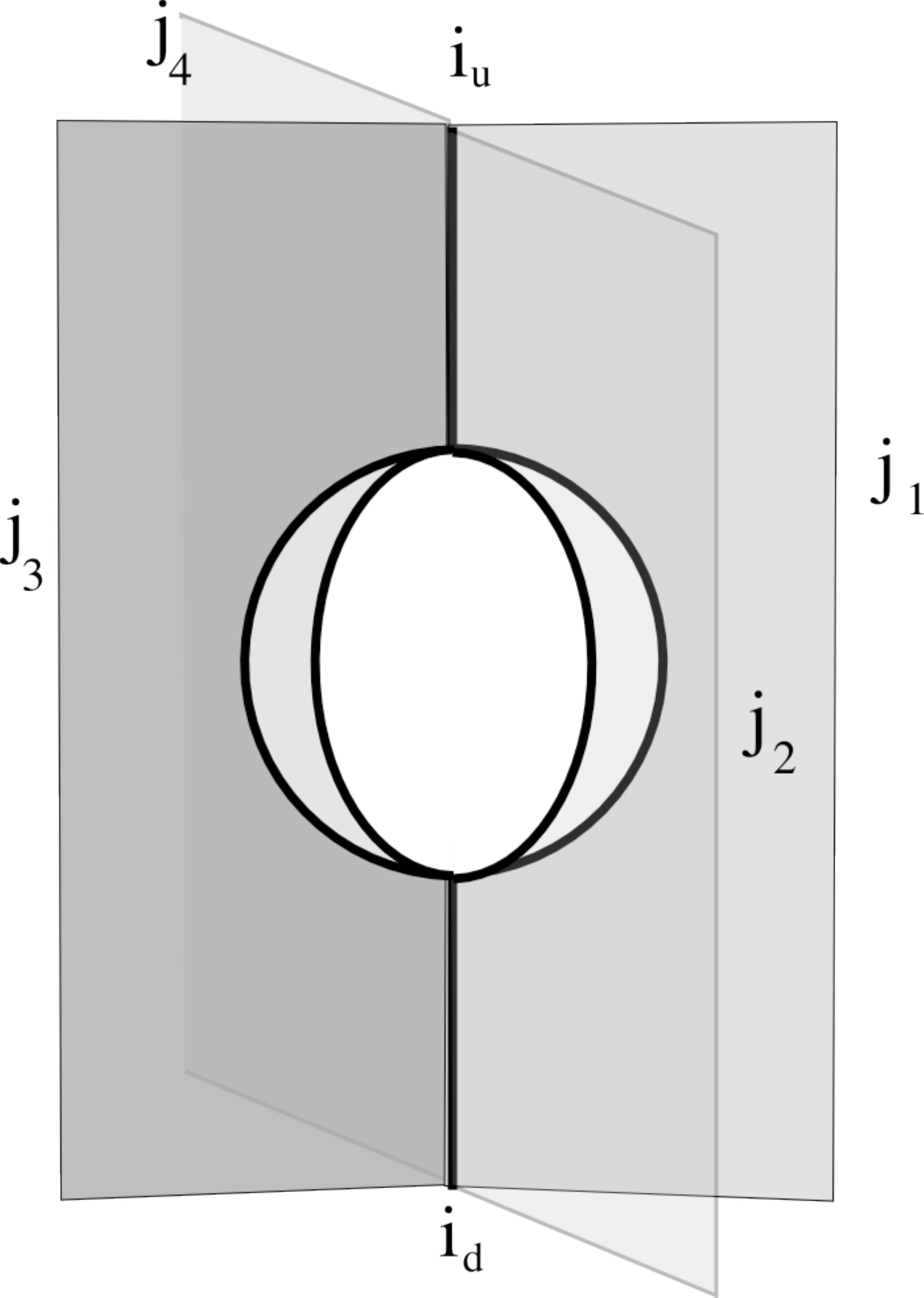}\hspace{8em}
\includegraphics[scale=0.15]{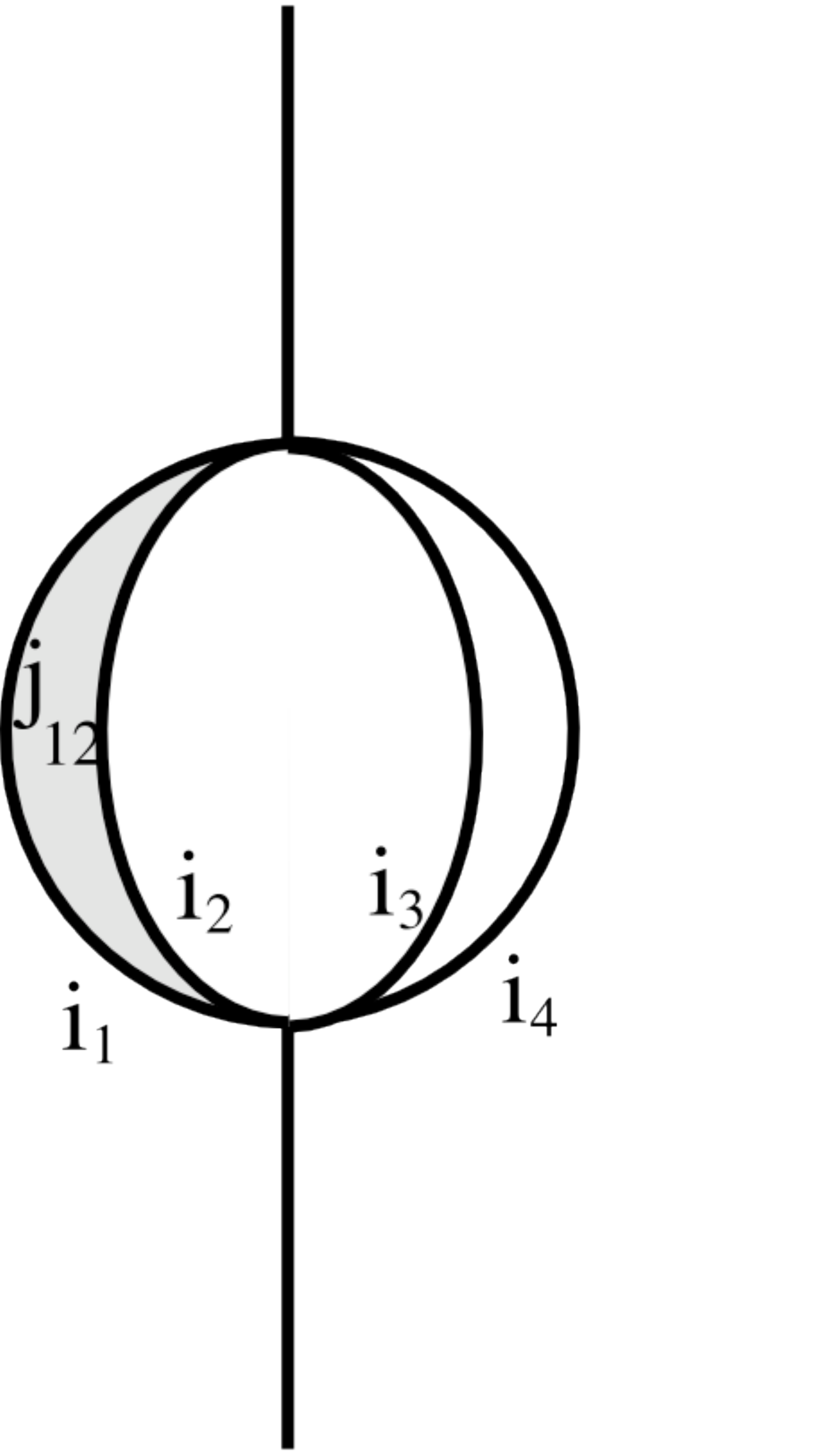}
\caption{The self-energy spinfoam. Left: boundary variables. Right: internal variables (only one of the six internal faces is highlighted).}
\end{figure}

Consider a triangulation $\Delta_2$ formed by two 4-simplices joined along four of their boundary tetrahedra. Its dual complex is illustrated in Figure 1.  The boundary triangulation $\partial\Delta_2$ if formed by two tetrahedra, joined by all their triangles.  The boundary graph is therefore formed by two nodes joined by four links.  Denote $i_u, i_d$ the intertwiners on the nodes and $j_a$, where $a=1,...,4$, the spins of the four links (see Figure 1). These are the boundary variables.  The internal variables are the four intertwiners $i_a$ on the four internal tetrahedra, and the six spins $j_{ab}$ on the six internal triangles (see Figure 2).
The amplitude is 
\be
W(j_a,i_u,i_d)=\lambda^2 \sum_{j_{ab},i_a} \prod_a\;d(j_a)\;\prod_{ab}\;d(j_{ab})\;
\; A_v(j_a,j_{ab},i_a,i_u) \; A_v(j_a,j_{ab},i_a,i_d).
\label{amplitude2}
\ee
We take $\lambda=1$ for simplicity in what follows.
To study if this amplitude diverges, it is convenient to set all external spins $j_a$ to zero.  (This is analogous
to the analysis of the vertex divergences in, say, $\lambda \phi^4$ QFT performed by setting the four external momenta to zero.)  In this case, $i_u=i_d=0$, where ``$0$" indicates the trivial intertwiners between trivial representations, and we have 
\be
W_2 \equiv W(0,0,0)=\sum_{j_{ab},i_a} \prod_{ab}\;d(j_{ab})\; \; A_v^2(0,j_{ab},i_a,0). 
\label{amplitude21}
\ee


\subsection{Simple example: $SU(2)$ BF}

\begin{figure}[t]
\centering
\includegraphics[scale=0.1]{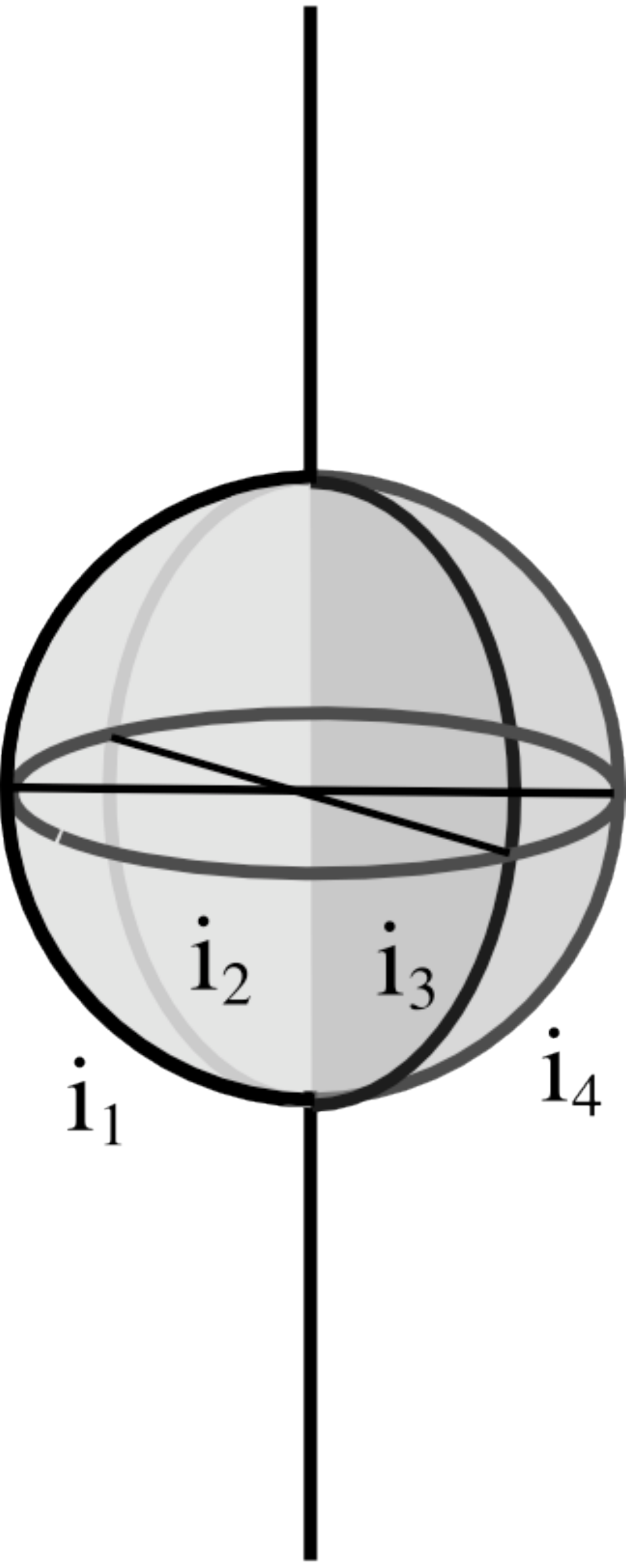}\hspace{8em}
\includegraphics[scale=0.13]{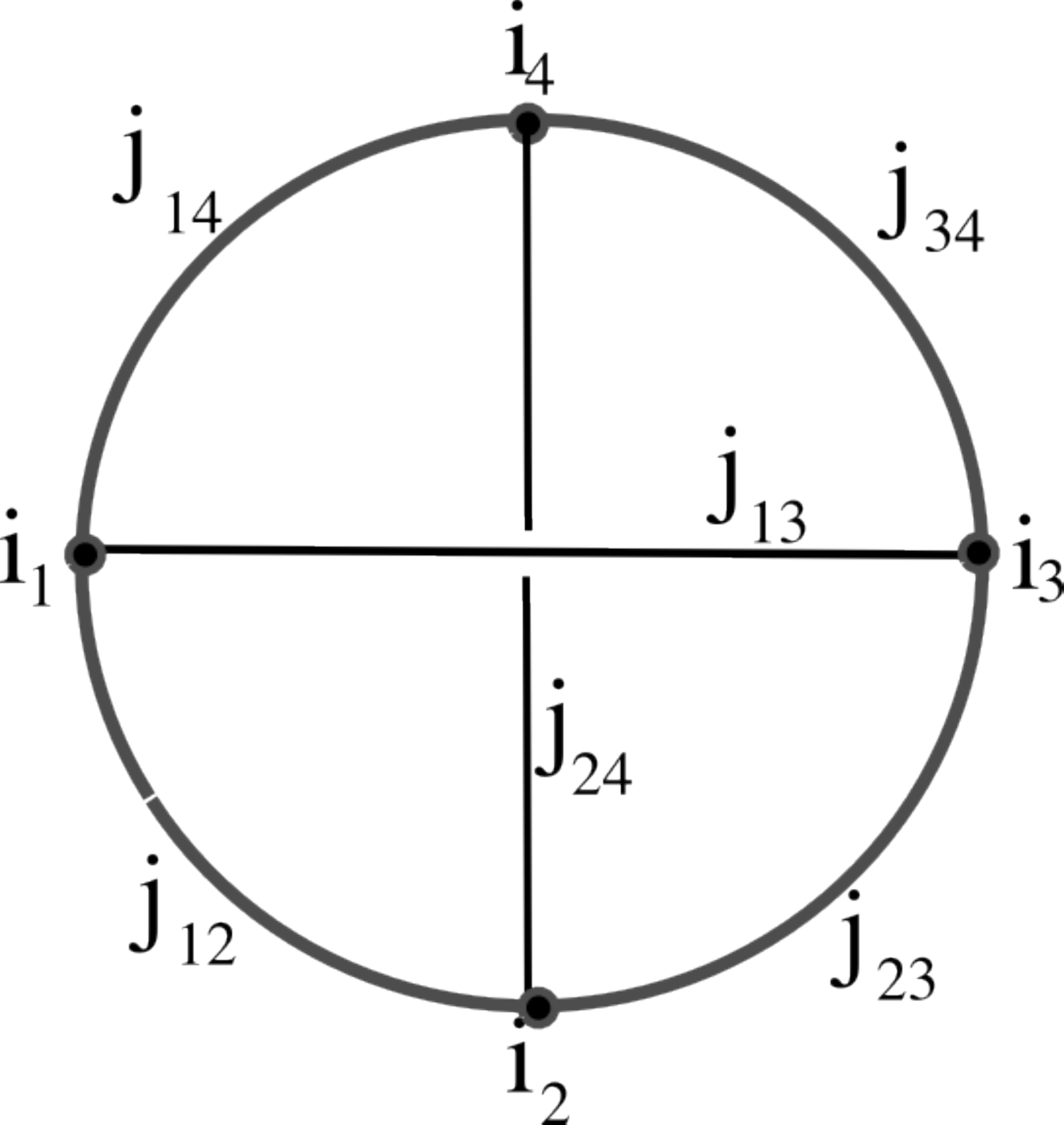}
\caption{The self-energy bubble. Left: the six faces. Right: equatorial section.}
\end{figure}

Before performing the computation for the gravity case, let us start with the simpler calculation for a (topological) four-dimensional $SU(2)$ BF theory.  In this case, 
\be
d(j)=(2j+1)^k
\ee
where the topological theory is obtained for $k=1$; and
\be
A_v(j_{nn'},i_n)=15j_N\! \left(j_{nn'}, i_n\right). 
\label{vertexBF}
\ee
When we set four spins sharing the same node to zero the intertwiners are all trivial, and the
normalized 15$j_N$ reduces to the 6$j$ symbol for the remaining six spins (see the Appendix.)
Thus (\ref{amplitude21}) becomes
\be
W_2=\sum_{j_{ab}} \prod_{ab}\; (2j_{ab}+1)^k\;
\left(6j \left(j_{ab}\right) \right)^2
\label{amplitude22BF}
\ee 
To estimate the degree of divergence of the above quantity we look at the large spin behaviour of the summand.  Using the Ponzano-Regge asymptotic expression for the 6-$j$ symbol \cite{PR}
\be
6j\sim \frac{1}{\sqrt{V}}\left(e^{iS}+e^{-iS} \right),
\ee
we have
\be\label{6j^2asympt}
(6j)^2 \sim \frac{1}{V}\left(2+e^{2iS}+e^{-2iS}\right)\sim \frac1V.
\ee
Here $V$ is the volume of a tetrahedron that has the $j_{ab}$ as lengths.  If $j_{ab}\sim j\to\infty$, then $V\sim j^3$ and $(6j)^2 \sim j^{-3}$. Combining with the measure factor gives
\be
W_2 \sim \sum_{j_{ab}} j^{6k-3}. 
\ee

The naive degree of divergence is given by the power of the summand plus  the number of 
sums. The Clebsch-Gordan conditions do not change this, because the volume of the region of ${j_{ab}<\Lambda}$ satisfying these relations still grows like $\Lambda^6$.  Hence
\be
D_{\rm BF}=6+6k-3.
\ee
The graph converges only if $k<-\frac12$. For the conventional $k=1$ it diverges as $\Lambda^9$, with $\Lambda$ an infrared cut off. 

For $k=1$, an alternative way to obtain the same estimate is the following. 
The expression \eqref{amplitude22BF} is precisely the Ponzano-Regge partition function of a triangulation formed by two tetrahedra glued by all their faces, and it is equivalent to
\be
W_2=\int dU_a \prod_{ab}\delta(U_aU_b^{-1}).
\ee
Integrating out the delta functions explicitly one by one, leaves
\be
W_2=(\delta(\mathbbm 1))^3.
\ee
Using the Plancherel expansion of the $\delta$ function (see \eqref{delta} in the Appendix) this becomes
\be
W_2=\Big(\sum_j\  (2j+1)\  {\rm tr}[R^{(j)}(\mathbbm 1)] \Big)^3= \Big(\sum_j\  (2j+1)^2\Big)^3
\ee
which indeed diverges as $\Lambda^9$, confirming the result above.   We now return to gravity.


\subsection{Gravity}
Inserting the gravitational vertex amplitude (\ref{vertex}) into equation (\ref{amplitude2}) gives
\be
W_2=\sum_{j_{ab},i_a} \prod_{ab}\;d(j_{ab})\; \; 
\left(\sum_{i_n^+,i^-_n} \,15j_N  \left(0,{\scriptstyle  \frac{1+\gamma}{2}}j_{ab}, i_n^+\right)\;15j_N \left(0,{\scriptstyle \frac{|1-\gamma|}{2}} j_{ab},  i_n^-\right) \prod_n f_{i_n^+,i^-_n}^{i_n}\right)^2.
\label{amplitude22}
\ee
Choosing the virtual spin $i$ in the pairing $((j_1,j_2),(j_3,j_4))$ to label the intertwiners,  fusion coefficients can be written in the form  \cite{CEE}
\begin{eqnarray}\label{fdefinition}
f^i_{i^+ i^-}(j_1,j_2,j_3,j_4)=\Big[(2i+1)(2i^++1)(2i^-+1)
\prod_{n=1}^4 (2j_n+1)\Big]^{\frac12}
\left\{\!\!
	\begin{array}{lll}
  j_{1}^+ &i^+ & j_{2}^+ \\[3pt]
  j_{1}^- &i^- & j_{2}^- \\[3pt]
  j_1 & i & j_{2}
  \end{array}
	\!\!\right\} 
\left\{\!\!
	\begin{array}{lll}
 j_{3}^+ &i^+ & j_{4}^+ \\[3pt]
  j_{3}^- &i^- & j_{4}^- \\[3pt]
  j_3 & i & j_{4}
  \end{array}
	\!\!\right\},
\end{eqnarray}
where the matrices are Wigner 9$j$-symbols (other choices of pairings are equivalent as $W_2$ does not depend on the basis chosen.) As above, the presence of four vanishing spins fixes the intertwiners and reduces the $15j_N$-symbol to the Wigner $6j$-symbol. One of the fusion coefficient (the one among all trivial representations) is equal to unity.  We have then
\be
W_2=\sum_{j_{ab},i_a} \prod_{ab}\;d(j_{ab})\;
\Big( 6j  \left({\scriptstyle  \frac{1+\gamma}{2}}j_{ab}\right) 
\  6j \left({\scriptstyle \frac{|1-\gamma|}{2}} j_{ab}\right) \prod_a  f_a \Big)^2
\label{amplitude23}
\ee
where $f_a$ stands for
\be
f_1=\sqrt{d(j_{12})d(j_{13})d(j_{14})}\left\{\!\!
	\begin{array}{ccc}
  j^+_{12} & j^+_{14} & j^+_{13} \\[3pt]
  j^-_{12} &j^-_{14}& j^-_{13} \\[3pt]
  j_{12} & j_{14} & j_{13}
  \end{array}
	\!\!\right\} 
	\label{f_1}
\ee
and so on cyclically. We show in Appendix B that for large $j_{ab}\simeq j$, the $f$'s behave like $j^{-3/4}$.  Together with the scaling (\ref{6j^2asympt}), we get
\be
W_2\sim \sum_{j_{ab}} (j^{k})^6 [(j^{-3/2})^2 (j^{-3/4})^4]^2.
\ee
We see that the degree of divergence with the fusion coefficients $f$ in (\ref{fusion}) is
\be
D_{W_2}=  6+6k-12.
\ee
Naively, the self-energy diverges as a power for $k=2$, and logarithmically for $k=1$.

Next, we consider the fusion coefficients normalized as in (\ref{fusion2}). As shown in the Appendix, these scale like $j^{-9/4}$. We then get
\be
\widetilde{W}_2\sim \sum_{j_{ab}} (j^{k})^6 [(j^{-3/2})^2 (j^{-9/4})^4]^2.
\ee
Therefore the degree of divergence with the fusion coefficients $\tilde f$ is
\be
D_{\widetilde{W}_2}=  6+ 6k-24, 
\ee
which converges for $k<3$. The stronger convergence could have been anticipated from (\ref{pippo}).


\newpage {\ }

\vskip1cm

\section{The ball}

\begin{figure}[t]
\centering
\includegraphics[scale=0.2]{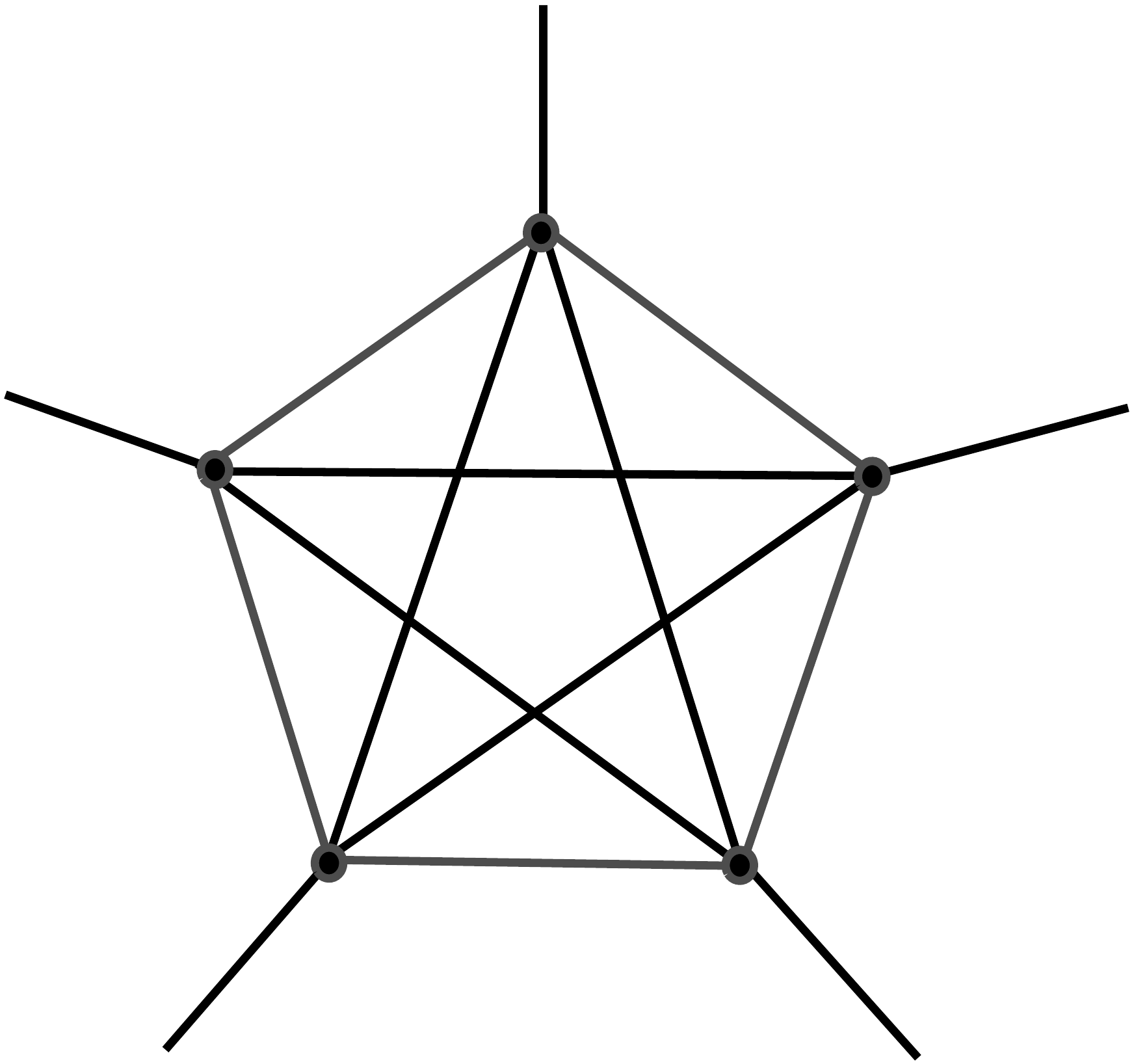}\hspace{8em}
\includegraphics[scale=0.15]{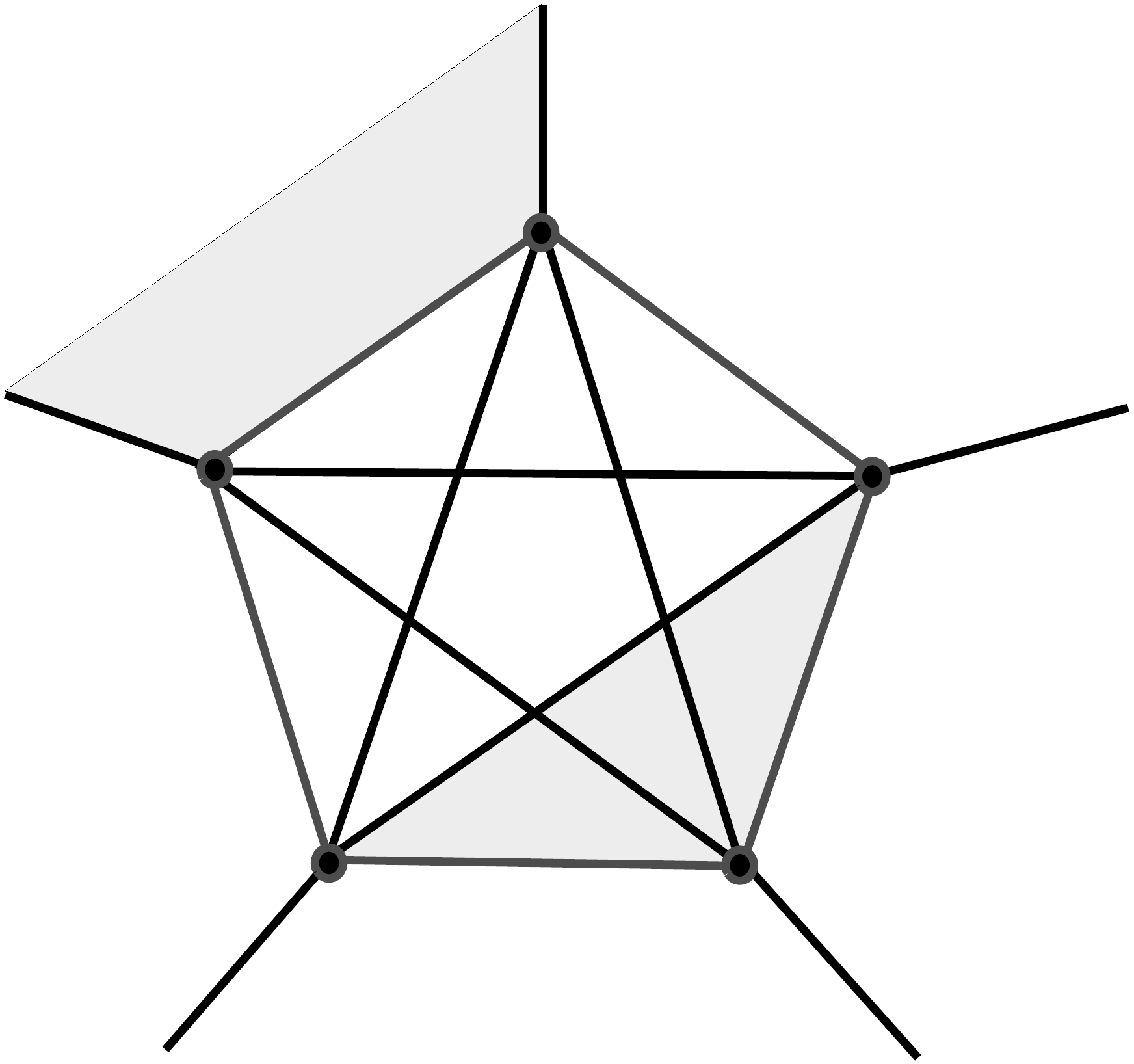}
\caption{The  ``ball'' vertex amplitude. Left: the five vertices and the ten+five edges. Right: one internal face and one external face are shown.}
\end{figure}

The next interesting diagram is a 4-simplex expanded with a 1-5 move. Such a graph
corresponds to a triangulation of a 3-ball with five 4-simplices, and boundary a 3-sphere triangulated with five tetrahedra. The divergence of this graph can be associated to the vertex renormalization.  In this case there are five vertices and the 
external data are ten spins and five tetrahedra. Let now $a,b,c=1,\ldots, 5$. To each couple $a<b$ is associated an internal intertwiner (tetrahedron); to each triple $a<b<c$ is associated an internal face (triangle shared by the three tetrahedra $(ab)$, $(bc)$ and $(ac)$). Putting the external spins and intertwiners to zero, the spinfoam amplitude is
\be
W_5=\lambda^5 \sum_{j_{abc}i_{ab}} \prod_{a<b<c} d(j_{abc}) \,A_v(0,j_{1ab},i_{1a},0)\ldots A_v(0,j_{ab5},i_{a5},0),
\ee
where we have momentarily reinserted $\lambda$ for clarity. 
As before, it reduces to an expression with $6j$ and $9j$-symbols, and the intertwiners are evaluated at fixed values: 
\be\label{W5completa}
W_5=\sum_{j_{abc}} \ \prod_{a<b<c} d(j_{abc}) \,6j\left({\scriptstyle  \frac{1+\gamma}{2}}j_{1ab}\right)\,
6j \left({\scriptstyle \frac{|1-\gamma|}{2}} j_{1ab}\right) \ldots 6j\left({\scriptstyle  \frac{1+\gamma}{2}}j_{ab5}\right)\,
6j \left({\scriptstyle \frac{|1-\gamma|}{2}} j_{ab5}\right)\prod_a  f_{1a} \ldots \prod_a f_{a5}.
\ee
Thus 
\be
W_5\sim\sum_{j_{abc}} j^{10k} [(j^{-3/2})^2 (j^{-3/4})^4]^5,
\ee
and the degree of divergence is
\be
D_{W_5}=10+10k-30.
\ee
The amplitude \eqref{W5completa} converges for $k=1$ and diverges logarithmically for $k=2$.  
With the $\tilde f$ fusion coefficients 
\be
\widetilde{W}_5\sim\sum_{j_{abc}} j^{10k} [(j^{-3/2})^2 (j^{-9/4})^4]^5
\ee
and the degree of divergence is
\be
D_{\widetilde{W}_5}=10+10k-60.
\ee
This converges for $k<5$.  

\newpage 

\section{Results}\label{results}

We have computed the naive degree of divergence of the first two interesting diagrams in loop quantum gravity,  self-energy and vertex, for different choices of normalizations and face amplitude.     With the fusion coefficients $f$ defined in \eqref{fusion}, the self-energy converges (naively) for $k<1$ and the vertex converges (naively) for $k<2$. Therefore with the $SO(4)$ measure factor (\ref{SO(4)}) the self-energy diverges as a power and the vertex diverges logarithmically, similar to what happens in QED.  With the $SU(2)$ measure factor (\ref{SU(2)}) the self-energy diverges logarithmically while the vertex is finite.  With the fusion coefficients $\tilde f$ defined in  \eqref{fusion2}, there are no divergences for $k<3$, therefore the two terms are finite with either of the two face amplitudes considered.  These results are summarized in Table 1.   

\vspace{.2cm}

\begin{table}[h]
\begin{center}
\begin{tabular}{|c|c|c|c|}
\hline
\ Fusion coefficients\ \ &Face amplitude& Self-energy & Vertex \\
\hline
$f$ &\ \ \ \ SU(2)($k=1$)\ \ \ \ &\ \ \ \  logarithmic \ \ \ \  & \ \ \ \  finite \ \ \ \  \\
 $f$&SO(4)($k=2$)& $\Lambda^6$ & \ \ \ \ logarithmic\ \ \ \   \\
 $\tilde f$ &\ \ \ \ SU(2)($k=1$)\ \ \ \ &finite & \ \ \ \  finite \ \ \ \  \\
 $\tilde f$ &SO(4)($k=2$)& finite & \ \ \ \  finite \ \ \ \   \\
\hline
\end{tabular}
\caption{Naive degree of divergence ($\Lambda$ is an infrared cut off).}
\end{center}
\label{default}
\end{table}
\vspace{-.3cm}

The degree of divergence depends strongly on the normalization of face and vertex amplitudes. 
It is thus crucial to have better control over these quantities, which are not uniquely fixed in the literature.

All these divergences are in the infrared. The conventional ultraviolet divergences of gravity \cite{weinberg76} do not appear in the theory; they are cut-off by the discretisation of the geometry given by the discrete character of the $j_f$ spins.  The divergences that appear have a clear geometrical interpretation: they correspond to  ``spikes" in the Regge triangulation, that is, very high ``pyramids" that can spike out from a spacetime region bounded by a finite sphere. 

We have not addressed the problem of characterizing \emph{all} divergent diagrams. 
In particular: if the ball and the bubble diverge, would a regularized bubble make the ball finite as well? How many are the diagrams that one needs to regularize to make \emph{all} of them finite?  These issues and the implications of these results for loop quantum gravity, will be discussed elsewhere. 

\acknowledgements
We are grateful to Elena Magliaro, Antonino Marcian\`o, Roberto Pereira, Daniele Pranzetti, Matteo Smerlak and Artem Starodubstev for numerous useful inputs and discussions.


\vskip2cm

\appendix
\section{Normalizations}

The Wigner 3j symbols
\be
i^{m_1m_2m_3}\equiv
\left(\!\begin{array}{ccc} j_{1}&j_{2}&j_{3} \\[2mm]  m_1&m_2&m_3  
\end{array}\!   \right)\!
\ee
are real trivalent intertwiners, namely invariant tensors in the tensor product $H_{j_1}\otimes H_{j_2} \otimes H_{j_3}$ of three $SU(2)$ representations, normalized by 
\be
\sum_{m_1m_2m_3}
\left(\!\begin{array}{ccc} j_{1}&j_{2}&j_{3} \\[2mm]  m_1&m_2&m_3  \end{array}\!   \right)\! 
\left(\!\begin{array}{ccc} j_{1}&j_{2}&j_{3} \\[2mm]  m_1&m_2&m_3  \end{array}\!   \right)\! 
=1.
\ee
They are related to the Clebsch-Gordan coefficients by a normalization factor
\be
\left(\!\begin{array}{ccc} j_{1}&j_{2}&J\\[2mm]  m_1&m_2&-M  \end{array}\!   \right)\!
=\frac{(-1)^{j_1-j_2-M}}{\sqrt{2J+1}}\  \langle j_1j_2, m_1m_2|JM \rangle.
\ee
The Wigner 3j symbols as usually represented by the vertex of three lines that join. 
\be
i^{abc} = \hspace{1em}
\begin{array}{c}\setlength{\unitlength}{1 pt}
\begin{picture}(50,48)          \put(0,4){$a$}
\put(17,40){$b$}     
     \put(35,4){$c$}
     \put(20,20){\line(-1,-1){10}}   
     \put(20,20){\line(1,-1){10}}         
      \put(20,20){\line(0,1){15}}       
      \put(20,20){\circle*{3}}
      \end{picture} 
\end{array}.
\ee
Joining free ends of these lines indicates contraction with the intertwiner  $g_{ab}$ between a representation and its dual. 

The 4-valent intertwiners can be written in terms of trivalent intertwiners. An orthonormal basis of 4-valent intertwiners is given by
\be
i_J^{abcd} \equiv \sqrt{2J+1} \ \sum_{mn}  g_{mn}
\left(\!\begin{array}{ccc} j_{1}&j_{2}&J\\[2mm]  a&b&m  \end{array}\!   \right)\!
\left(\!\begin{array}{ccc}J& j_{3}&j_{4}\\[2mm]  n&c&d  \end{array}\!   \right) \equiv  \sqrt{2J+1}\  i^{abm}i_m{}^{cd}.
\ee
Orthonormality can be proven explicitly by writing 
\begin{eqnarray}\nonumber
i_J^{abcd}i_K{}_{abcd}\!\! &=&\!\! \sqrt{2J+1}\sqrt{2K+1}\ g_{aa'}g_{bb'}g_{cc'}g_{dd'} g_{mn}
 g_{m'\!n'}\!
\left(\!\begin{array}{ccc} j_{1}&j_{2}&J\\[2mm]  a&b&m  \end{array}\!   \right)\!
\left(\!\begin{array}{ccc}J& j_{3}&j_{4}\\[2mm]  n&c&d  \end{array}\!   \right)\!
\left(\!\begin{array}{ccc} j_{1}&j_{2}&K\\[2mm]  a'&b'&m'  \end{array}\!   \right)\!
\left(\!\begin{array}{ccc}K& j_{3}&j_{4}\\[2mm]  n'&c'&d'  \end{array}\!   \right) \\
&=&
 \sqrt{2J+1}\sqrt{2K+1}\  \delta_{JK}\frac{\delta^{m}_n\delta^{n}_m}{(2J+1)(2K+1)}=\delta_{JK}.
\end{eqnarray}
The standard graphical notation for the 4-valent intertwiners 
indicates  $i^{abm}i_m{}^{cd}$, and not the normalized intertwiners $i_J^{abcd}= \sqrt{2J+1}\  i^{abm}i_m{}^{cd}$, that is
 \be
i_j^{abcd} = \sqrt{2j+1}\hspace{1em}
\begin{array}{c}\setlength{\unitlength}{1 pt}
\begin{picture}(50,40)          \put(0,4){$a$}\put(0,30){$b$}          \put(45,4){$d$}\put(45,30){$c$}    \put(10,10){\line(1,1){10}}\put(10,30){\line(1,-1){10}}    \put(30,20){\line(1,1){10}}\put(30,20){\line(1,-1){10}}          \put(20,20){\line(1,0){10}}\put(22,25){$j$}          \put(20,20){\circle*{3}}\put(30,20){\circle*{3}}\end{picture} 
\end{array}\,.
\ee

Since the last form an orthonormal basis on the invariant part of the product of four representation spaces, we have 
\be
\sum_J  \ i_J^{abcd}\ i_J^{a'b'c'd'}
=
\sum_J \ i_J^{acbd}\ i_J^{a'c'b'd'}
\ee
from which the recoupling theorem follows easily.  In terms of these quantities, and the unitary representation matrices of SU(2), $R^{(j)}{}^a{}_{b}(U)$, we have
\begin{eqnarray}
\int dU\ R^{(j)}{}^a{}_{b}(U)\ R^{(k)}{}^c{}_{d}(U)&=&\frac1{2j+1}\ \delta^{jk}g^{ac}g_{bd}\\  
\int dU\ R^{(j)}{}^a{}_{d}(U)\ R^{(k)}{}^b{}_{e}(U)\ R^{(l)}{}^{c}{}_{f}(U)&=&
i^{abc}i_{def}
\\
\int dU\ R^{(j)}{}^a{}_{e}(U)\ R^{(k)}{}^b{}_{f}(U)\ R^{(l)}{}^{c}{}_{g}(U)R^{(m)}{}^d{}_{h}(U)\ &=&
\sum_J\  i_J^{abcd}i_J{}_{efgh}
\\
\end{eqnarray}
and
\be
\delta(U) = \sum_j\  (2j+1)\  {\rm tr}[R^{(j)}(U)] 
\label{delta}
\ee
which are the basic formulas for deriving the spinfoam representation of BF theory.   Expanding the BF partition function
\be
Z= \int dU_l \prod_f\delta(U_{f_1}...U_{f_n})
\ee
in representations and using the formulas above we obtain the well known Ponzano-Regge expression 
\be
Z= \sum_{j_f} \prod_f (2j_f+1) \prod_v 6j(j_f)
\ee
in the 3d case. While in the (SU(2)) 4d case we obtain 
\be
Z= \sum_{j_f, i_e} \prod_f (2j_f+1) \prod_v A(j_f, i_e), 
\ee
where the vertex amplitude is the contraction of five \emph{normalized} intertwiners.
\be
 A(j_f, i_e)=\otimes_a i_a. 
\ee
This is not the 15j symbol as usually defined in representation theory texts:
\be
 15j(j_{ab}, j_a)\equiv\left\{\!\!
	\begin{array}{ccccc}
  j_{12} &j_{23}  & j_{34}& j_{45}& j_{51} \\[3pt]
  j_{35} & j_{41} & j_{52}&j_{13}&j_{25} \\[3pt]
  j_1 & j_2 & j_3 & j_4 & j_5
  \end{array}
	\!\!\right\}
\ee
but rather 
\be
 A(j_{ab}, j_a)=\ 15j_N\!(j_{ab}, j_a)\equiv\sqrt{\prod_a (2j_a+1)}\ 15j(j_{ab}, j_a). 
\label{15jN}
\ee

\vskip1cm


\section{Asymptotics of fusion coefficients with a single zero spin}

When one of the four spins is zero, say $j_4$, the fusion coefficients reduce, up to a sign, to
\begin{equation}\label{fblob}
\tilde f^i_{i^+ i^-}(j_1,j_2,j_3,0)=\pm\delta_{i^+,j_{3}^+}\delta_{i^-,j_{3}^-}\delta_{i,j_3}
\  9j(j_1,j_2,j_3)\;,
\end{equation}
where
\be
9j(j_1,j_2,j_3)\equiv\left\{\!\!
	\begin{array}{ccc}
  j_{1}^+ &j_{3}^+  & j_{2}^+ \\[3pt]
  j_{1}^- & {j_{3}^-} & j_{2}^- \\[3pt]
  j_1 & j_3 & j_{j2}
  \end{array}
	\!\!\right\} 
	\ee
is a Wigner 9$j$ symbol. 
We are interested in the asymptotic behavior of  \eqref{fblob} for large spins. (For a numerical
approach, see \cite{Igor}.)
The $9j$-symbol with two degenerate columns can be written as \cite{Wigner}
\begin{eqnarray}\label{degenerate9j}
\left\{\!\!
	\begin{array}{ccc}
  a & f & c \\[3pt]
  b & g & d \\[3pt]
  a+b & h & c+d
  \end{array}
	\!\!\right\} =&\;\; (-1)^{f-g+a+b-(c+d)} \;\;\left(\!\!
	\begin{array}{ccc}
  f & g & h \\[3pt]
  a-c & b-d &  -(a+b-(c+d))
  \end{array}
	\!\!\right)\;\;\times
	\\[10pt]
	&\hspace{-9em} \times\;\; \sqrt{\frac{(2a)!(2b)!(2c)!(2d)!(a+b+c+d-h)!(a+b+c+d+h+1)!}{(2a+2b+1)!(2c+2d+1)! (a+c-f)!(a+c+f+1)! (b+d-g)! (b+d+g+1)!}} \;.
\end{eqnarray}
The degenerate $3j$-symbol, in which the third spin is the sum of the first two can be derived
easily from well-known expressions for the 3$j$ symbol \cite{Wigner}:
\begin{equation}\label{degenerate3j}
\left(\!\!
\begin{array}{ccc}
 f & g & f+g \\[3pt]
  m & n & -(m+n)
\end{array}
\!\!\right)
=(-1)^{f-g+m+n}\sqrt{\frac{(2f)!(2g)!(f+g-m-n)!(f+g+m+n)!}{(1+2f+2g)!(f-m)!(f+m)!(g-n)!(g+n)!}}.
\end{equation}
We can use the representations \eqref{degenerate9j} and \eqref{degenerate3j} both for $0\leq\gamma<1$ and for $\gamma>1$.  This is because in the first case we have $j^+ +j^-=j$; while in the second have $j+j^-=j^+$: in either case there is one entry which is given by the sum of the others.   Both in \eqref{degenerate9j} and \eqref{degenerate3j} some factorials are written in the form $(1+x)!$; we substitute them with the expression $(1+x)x!$\,. Then, as we are interested in the large spin behavior, we apply Stirling's approximation
$$x!\sim\sqrt{2\pi x}\;e^{-x+x\log x}$$
to all factorials in the resulting expression. All factors coming from the exponential in the Stirling formula cancel out, giving 1. The only contribution is provided by the monomials $(1+x)$ we factored out, and from $\sqrt{2\pi x}$ in Stirling's expression. So we obtain easily the asymptotic expression
\begin{eqnarray}
\label{fspin0as}
9j(j_1,j_2,j_3)\sim\pm\,\beta_{\gamma}\frac{\sqrt{2}}{\pi^{1/4}}\left[
\frac{1}{\sqrt{j_1 \,j_2\,j_3\,(j_1+j_2-j_3)(j_3+j_1-j_2)(j_3+j_2-j_1)(j_1+j_2+j_3)^3}}\right]^{1/2}
\end{eqnarray}
where
\be
\beta_\gamma=\begin{cases}\big(\frac{1}{1-\gamma^2}\big)^{1/4}&0\leq\gamma<1\\[6pt]
         \big(\frac{1}{\gamma^2-1}\big)^{3/4}&\gamma>1\end{cases}
\ee
Therefore the 9$j$ Wigner symbols that appear in the fusion coefficients behave like $j^{-9/4}$ for large spins.  Accordingly the fusion coefficients behave as
\be
 f \sim \frac{1}{j^{3/4}}, \hspace{4em}
\tilde f \sim \frac{1}{j^{9/4}}. 
\ee
These are the asymptotic expressions used in the text.


\begin{thebibliography}{99}

\bibitem{EPR}  J.\,Engle, R.\,Pereira, C.\,Rovelli: 
``The loop-quantum-gravity vertex-amplitude", 
Phys. Rev. Lett., 99 (2007) 161301;  arXiv:0705.2388.

\bibitem{EPR2}  J.\,Engle, R.\,Pereira, C.\,Rovelli: 
``Flipped spinfoam vertex and loop gravity", 
Nucl. Phys. B798 (2008) 251-290;   arXiv:0708.1236.\\
R.\,Pereira, ``Lorentzian LQG vertex amplitude",
Class. Quant. Grav. 25 (2008) 085013;  arXiv:0710.5043. 

\bibitem{FK}  
L.\,Freidel, K.\,Krasnov, 
``A New Spin Foam Model for 4d Gravity".
Class. Quant. Grav. 25 (2008) 125018, 2008;   arXiv:0708.1595.

\bibitem{LS}
E.R.\,Livine, S.\,Speziale, 
``A New spinfoam vertex for quantum gravity", 
Phys. Rev. D76 (2007) 084028;  arXiv:0705.0674. 
\\
E.R.\,Livine, S.\,Speziale, 
  ``Consistently Solving the Simplicity Constraints for Spinfoam Quantum Gravity,''
  Europhys.\ Lett.\  81  (2008) 50004;arXiv:0708.1915. 

\bibitem{EPRL}  J.\,Engle, E.R.\,Livine, R.\,Pereira, C.\,Rovelli:  
``LQG vertex with finite Immirzi parameter"
Nucl. Phys. B799 (2008) 136-149; arXiv:0711.0146.

\bibitem{book} C.\,Rovelli, {\em Quantum Gravity} (Cambridge University Press, Cambridge 2004). \\ 
T.\,Thiemann, {\em Modern canonical quantum general relativity}, (Cambridge
  University Press, Cambridge, 2007). 
    
\bibitem{lqg2}  
C.\,Rovelli, L.\,Smolin,
 ``Knot theory and quantum gravity'' 
{Phys.\,Rev.\,Lett} {61} (1988) 1155-1158. \\
C.\,Rovelli, L.\,Smolin,
``Loop space representation for quantum general relativity", 
{Nucl.\,Phys.} {B331}  (1990) 80-152.\\
 A.\,Ashtekar, C.\,Rovelli, L.\,Smolin,
 ``Weaving a classical geometry with quantum threads",
{Phys.\,Rev.\ Lett}  {69}  (1992) 237.\\
C.\,Rovelli,
``A generally covariant quantum-field theory and a preiction on quantum measurements of geometry", {Nucl.\,Phys.} {B405}  (1993) 797-815.\\
C.\,Rovelli, L.\,Smolin,
 ``Discreteness of Area and Volume in Quantum Gravity", 
{Nucl.\,Phys.} {B442} (1995)  593-619;
{Nucl.\,Phys.} {B456}  (1995) 734.

\bibitem{Wilson} K.\,Wilson, ``Confinement of Quarks", Phys. Rev. D10 (1974) 2445.


\bibitem{misner} C.\,Misner, ``Feynman quantization of general
relativity", {Rev.\,Mod.\,Phys.\,} {29} (1957) 497.\\
S.\,W.\,Hawking, ``The Path-Integral Approach to Quantum
Gravity'', in {\sl General Relativity: An Einstein Centenary Survey},
S.\,W.\,Hawking and W.\,Israel eds (Cambridge University Press, Cambridge
1979).

\bibitem{Regge} 
T.\,Regge, ``General Relativity Without Coordinates",
Nuovo Cimento 19 (1961) 558-571.

\bibitem{Perez:2000bf}
  A.\,Perez,
  ``Finiteness of a spinfoam model for Euclidean quantum general  relativity,''
  Nucl. Phys.  B{599}  (2001) 427.

\bibitem{AC}  A.\,Perez, C.\,Rovelli: ``A spin foam 
model without bubble divergences",
Nucl. Phys.   B599 (2001) 255-282; gr-qc/0006107. 

\bibitem{C1} L.\,Crane, A.\,Perez, C.\,Rovelli: 
``Finiteness in spinfoam quantum gravity", 
Phys. Rev. Lett. 87 (2001) 181301.

\bibitem{C2} L.\,Crane, A.\,Perez, C.\,Rovelli:
``A finiteness proof for the Lorentzian state sum 
spinfoam model for quantum general relativity", 
gr-qc/0101088. 

\bibitem{semiclassical} 
E.\,Magliaro, C.\,Perini, C.\,Rovelli, ``Numerical indications on the semiclassical limit of the
flipped vertexÓ, Class. Quant. Grav. 25 (2008) 095009, arXiv:0710.5034.\\
B.\,Dittrich, S.\,Speziale, ``Area-angle variables for general relativity,''
  New J.\,Phys.\,10 (2008) 083006; arXiv:0802.0864.\\
E.\,Alesci, E.\,Bianchi, E.\,Magliaro, C.\,Perini, ``Intertwiner dynamics in the flipped vertexÓ, 
arXiv:0808.1971.\\
E.\,Bianchi,  A.\,Satz, ``Semiclassical regime of Regge calculus and spin foamsÓ, 
arXiv:0808.1107.\\
F.\,Conrady, L.\,Freidel, ``On the semiclassical limit of 4d spin foam modelsÓ, 
arXiv:0809.2280.

\bibitem{Reisenberger:2000zc}
  M.P.\,Reisenberger, C.\,Rovelli,
  ``Spacetime as a Feynman diagram: The connection formulation,''
  Class.\ Quant.\ Grav.\  {18} (2001) 121.
  
\bibitem{GFT} 
  D.\,Oriti,
  ``The group field theory approach to quantum gravity,''
in  {\em Approaches to Quantum Gravity - toward a new understanding of space, 
time, and matter}, D.\,Oriti editor (Cambridge University Press 2008).
  arXiv:gr-qc/0607032. \\
L.\,Freidel,  ``Group Field Theory: an overview'',
    {Int. Journ. Theor. Phys.} {44} (2005) 1769-1783.\\ 
    R.\,DePietri, L.\,Freidel, K.\,Krasnov, C.\,Rovelli ``Barrett-Crane
    model from a Boulatov-Ooguri field theory over a homogeneous space",
    {Nucl. Phys.} {B574} (2000) 785-806. 

\bibitem{PR} 
G.\,Ponzano, T.\,Regge, ``Semiclassical limit of Racah coefficientsÓ. in {\em Spectroscopic and Group Theoretical Methods in Physics} F.\,Block editor (North Holland, Amsterdam, 1968).

\bibitem{weinberg76} S.\,Weinberg, ``Ultraviolet divergences in quantum
theories of gravitation", in {\sl General Relativity: An Einstein
Centenary Survey}, S.\,W.\,Hawking and W.\,Israel eds (Cambridge University
Press, Cambridge 1979).

\bibitem{CEE} E.\,Alesci, E.\,Bianchi, E.\,Magliaro, C.\,Perini, 
``Asymptotics of LQG fusion coefficients",  arXiv:0809.3718.

\bibitem{Igor} I.\,Khavkine, ``Evaluation of new spin foam vertex amplitudes,Ó arXiv:0809.3190.

\bibitem{Wigner} A.\,Yutsin, I.\,Levinson, V.\,Vanagas, {\em Mathematical Apparatus of the Theory of Angular Momentum} (Israel Program for Scientific Translation, Jerusalem, 1962).\\
D.\,Varshalovich, A.N.\,Moskalev, and V.K.\,Khersonskii, {\em Quantum Theory of Angular
Momentum} (World Scientific, Singapore 1988).

\end{thebibliography}
\end{document}